\documentclass[11pt,twoside,a4paper]{article}
\usepackage{srcltx}
\usepackage{amssymb,amsmath}
\usepackage{amsfonts}
\usepackage{indentfirst}
\usepackage{graphicx}
\newtheorem{corollary}{Corollary}
\newtheorem{theorem}{Theorem}

\newtheorem{definition}{Definition}

\begin{document}
\title{Elementary symmetric functions of two solvents of a quadratic matrix equation}

\author{\normalsize M A
Jivulescu$^{1,2}$, A Napoli$^1$, A Messina$^1$\\
\small ${}^1$ MIUR, CNISM and Dipartimento di Scienze Fisiche ed
Astronomiche, \\\small Universit\`{a} di Palermo, via Archirafi
36, 90123 Palermo, Italy\\\small ${}^2$  Department of
Mathematics, " Politehnica" University of Timi\c{s}oara,
\\\small P-ta Victoriei Nr. 2,
 300006 Timi\c{s}oara,
Romania\\
\small {\it Email Address:$^{1,2}$ maria.jivulescu@mat.upt.ro}\\
\small {\it Email Address:$^1$ messina@fisica.unipa.it}}

\maketitle

\begin{abstract}
Quadratic matrix equations occur in a variety of applications. In
this paper we introduce new permutationally invariant functions of
two solvents of the $n\times n$ quadratic matrix equation
$X^2-\mathcal{L}_1X-\mathcal{L}_0=0$, playing the role of the two
elementary symmetric functions of the two roots of a quadratic
scalar equation. Our results rely on the  connection existing
between the QME and the theory of linear second order difference
equations with noncommutative coefficients. An application of our
results to a simple physical problem is briefly discussed.

\end{abstract}

\medskip


\section{INTRODUCTION} \vspace{0.5cm}

Matrix language  belongs to all Sciences  where multicomponent
variables and noncommutativity enter scene in the description of
the system under scrutiny. In such situations matrix methods allow
indeed compact formulations and elegant resolutions of linear and
sometimes nonlinear problems. As an example, consider the time
development of a quantum system investigated in the Heisenberg
picture. One is in general faced with a matrix, often nonlinear,
systems of differential equations typically very difficult to
handle due to the noncommutativity of the involved observables.
Even when the problem under investigation enables the decoupling
of this system achieving linear differential Heisenberg equations
in a $n$-dimensional Hilbert space, the application of the
characteristic equation method, so useful to write down the
general integral in the scalar case, originates a $n$-dimensional
matrix nonlinear algebraic equation whose resolution  cannot
unfortunately rely on general propositions. Differently from the
case $n=1$ (that is the scalar case), the fundamental theorem of
algebra does not indeed hold, so that the roots (usually called
solvents) of  an algebraic matrix equation may exist or not and
even stipulating their existence, their number cannot be simply
related to the degree of the equation and could in particular be
infinite. It is thus not surprising that the theory of algebraic
matrix equations both for its wide applicability  and as research
subject aiming  extending  the beautiful  chapter of the scalar
algebraic equations, has recently received a great deal of
attention in the mathematical literature (Gantmacher 1998, Gohberg
2006, Horn 1999). Here we recall the analysis of the spectral
properties of the associated
 matrix pencil via that of the  matrix equation
itself (Gohberg et al. 1978, Krein et al. 1978). Many research
papers on the properties of a quadratic matrix equation,
especially numerical approaches, have appeared over the last
years (Bai et al. 2005, Butler et all 1985, Dennis et all 1978,
Highham 1987, 2000, 2001, Shurbet et al 1974).
\par Our paper investigates
the second degree $n$-dimensional matrix algebraic equations
possessing the following canonical form
\begin{equation}\label{QME}X^2-\mathcal{L}_1X-\mathcal{L}_0=0\end{equation}
where the unknown $X$ and the two, generally not commuting,
coefficients $\mathcal{L}_0,\mathcal{L}_1$ belong to
$M_n(\mathbb{C})$, the algebra of all complex square matrices of
order $n$. We call this equation a right quadratic matrix equation
(RQME). Its resolution is far from being trivial essentially
because the simple resolutive formula holding for the quadratic
scalar equation (QSE) is not in general applicable due to the
noncommutative character of the problem.\\ Here \textit{right}
means that the linear term in eq. (\ref{QME}) has the form
$\mathcal{L}_1 X$ instead of $X\mathcal{L}_1$ or
$\mathcal{L}_1X+X\mathcal{L}'_1$ defining a \textit{left} (LQME)
or a \textit{bilateral} quadratic matrix equation (BQME)
respectively. Right, left and bilateral matrix equations are
particular cases of  the Riccati algebraic matrix equation (Hore
et al. 1999) whose canonical form is
\begin{equation}XAX+BX+XC+D=0\end{equation}
In the next section we will show that the resolution of this
equation is always traceable back to that of a related RQME.
\par Thus, in this paper we concentrate on the properties possessed by eq. (\ref{QME}) and explore whether or at which extent
they may be thought of as generalizations of well-known properties
of elementary QSE. In this spirit we seek analogies and
differences between the two quadratic equations, matrix and
scalar, and, in connection with eq. (\ref{QME}),  we  formulate
the following questions:
\par \textit{Q1) To find the number of its solvents;}
\par \textit{Q2) To express its solvents, if any, in terms of its
coefficients and viceversa;}
\par \textit{Q3) To define  symmetric functions of a pair of solvents.}
\\It is well known that all three questions may be satisfactorily
coped with when the quadratic  equation is the scalar one. We
shall show that the matrix nature of the unknown  $X$ of eq.
(\ref{QME}) together or not with the noncommutativity between the
two coefficients as well as between a solvent and the same
$\mathcal{L}_1$ and $\mathcal{L}_0$, determine properties of
eq.(\ref{QME}) with no counterpart among those possessed by the
scalar equation. Our investigation will bring to light interesting
relations between the solvents of eq. (\ref{QME})  and its
coefficients $\mathcal{L}_1$ and $\mathcal{L}_0$ interpretable as
generalizations of the classical Girard-Newton and Waring formulas
(Sansone 1952).

In the  section 3, devoted to the questions \textit{Q1)} and
\textit{Q2)}, we introduce the notation,  give a short sketch of
the existing literature and report ad-hoc built examples
supporting some theoretical  statements. Globally speaking,
develop step by step convincing arguments enabling us to
introduce, on a heuristic basis, our new definition of elementary
functions of two solvents of a quadratic matrix equation.

The main and novel results of this paper are constructed in
connection with our reply to \textit{Q3)} and are reported in
section 4 where we  introduce and study the properties of the
permutationally invariant functions of two solvents of the
quadratic matrix equation.
  \vspace{0.2cm}
\section{Mapping a Riccati algebraic equation into a RQME}
\par The Riccati
algebraic matrix equation (Hore et al. 1999) in the unknown $Z\in
M_n(\mathbb{C})$
\begin{equation}ZAZ+\widetilde{\widetilde{\mathcal{L}_1}}Z+Z\widetilde{\widetilde{\mathcal{L'}_1}}+\widetilde{\widetilde{\mathcal{L}_0}}=0, \quad\quad \widetilde{\widetilde{\mathcal{L'}_1}}, \widetilde{\widetilde{\mathcal{L}_1}},\widetilde{\widetilde{\mathcal{L}_0}}\in M_n(\mathbb{C})\end{equation}
assuming the existence of $A^{-1}$ and introducing the change of
variable $Z=YA^{-1}$ may be reduced to the following BQME in $Y$:
\begin{equation}\label{BQME}Y^2+\widetilde{\mathcal{L}_1}Y+Y\widetilde{\mathcal{L'}_1}+\widetilde{\mathcal{L}_0}=0\end{equation}
with
$\widetilde{\mathcal{L}_1}=\widetilde{\widetilde{\mathcal{L}_1}}$
and
$\widetilde{\mathcal{L'}_1}=A^{-1}\widetilde{\widetilde{\mathcal{L'}_1}}A$
and
$\widetilde{\mathcal{L}_0}=\widetilde{\widetilde{\mathcal{L}_0}}A$.
  It is worth noticing that the BQME (\ref{BQME}), in its most general case, that is when both
$\widetilde{\mathcal{L}}_1\neq0$ and
$\widetilde{\mathcal{L'}}_1\neq0$, is  traceable back to the RQME
 in the canonical form (\ref{QME})  by furtherly changing the variable
$Y$ to $X=-\widetilde{\mathcal{L'}}_1-Y$, where now
$\mathcal{L}_1=\widetilde{\mathcal{L}_1}-\widetilde{\mathcal{L}'_1}$
and
$\mathcal{L}_0=\widetilde{\mathcal{L}_1}\widetilde{\mathcal{L}_1'}-\widetilde{\mathcal{L}_0}$.
The particular case when $\widetilde{\mathcal{L}}_1=0$ and
$\widetilde{\mathcal{L'}}_1\neq0$ transforms the BQME (\ref{BQME})
into a LQME which, by applying the above proposed change of
variable, may be traced back to the RQME (\ref{QME}) with
$\mathcal{L}_1=-\widetilde{\mathcal{L}_1'}$ and
$\mathcal{L}_0=-\widetilde{\mathcal{L}_0}$.
\par In all the  cases above considered the original matrix equation and its related RQME equation are simultaneously possible
or impossible and when the two sets of solvents are not empty, a
biunivocal correspondence may be established. \\ It is worth
noting, in passing, that in the symmetric case of BQME (
$\widetilde{\mathcal{L}_1}=\widetilde{\mathcal{L}_1'}$), called
the \textit{symmetrical bilateral } quadratic matrix equation (
SBQME), the related right equation becomes
$X^2=\widetilde{\mathcal{L}}_1^2-\widetilde{\mathcal{L}_0}$, whose
solutions
 exist if  the square root of  $\widetilde{\mathcal{L}}_1^2-\widetilde{\mathcal{L}_0}$ exists and are denoted by the matrix multivalued symbol
$\sqrt{\widetilde{\mathcal{L}}_1^2-\widetilde{\mathcal{L}_0}}$.\\
The systematic possibility of transforming LQME, BQME and Riccati
algebraic matrix equations into a RQME justify why in this paper
we concentrate on the canonical form (\ref{QME}) only, thereafter
simply called QME.

\section{ Matrix versus scalar quadratic equation }
\par In this section we give our reply to the questions  \textit{Q1)} and \textit{Q2)} beginning
 a comparison between
a QSE with a QME.\vspace{0.2cm} \subsection{\textbf{On the number
of solvents}} \vspace{0.2cm} \par The fundamental theorem of
algebra due to D'Alembert states that each scalar algebraic
equation of degree $n$ in the field $\mathbb{C}$ always admits at
least one root in $\mathbb{C}$. This theorem, together with the
well-known Ruffini's reminder-factor theorem, implies that the
number of solutions of the equation is exactly equal to $n$. The
examples reported  in Appendix A clearly demonstrate that the
fundamental theorem of algebra does not hold for a matrix equation
which, in turn, may be impossible, possible with finitely-many or
infinitely-many solvents in $M_n(\mathbb{C})$.

It is easy to persuade oneself that analogous examples for matrix
algebraic equations of higher degree may be given. Thus, we do not
have at our disposal a general rule directly involving
$\mathcal{L}_0$ and $\mathcal{L}_1$ to assess conclusions on the
number of solvents of a matrix algebraic equation without solving
it.
\par It is of relevance to underline at this point that a
resolutive algorithm of a QME does  exist(Gantmacher 1998) and
stems from a theorem stating that if $X$ is any solution of eq.
\eqref{QME} then $g(X)=0$, where $g(\lambda)\equiv \det
Q(\lambda)=\det (\lambda^2E-\mathcal{L}_1\lambda-\mathcal{L}_0)$
is a polynomial in $\lambda$ of degree $2n$. The key for the
resolution of eq.\eqref{QME} is to exploit the knowledge of the
roots of $g(\lambda)$ associated to eq.\eqref{QME} and their
respective multiplicities to build an appropriate change of the
variable $X$ into eq.\eqref{QME} which has the merit of reducing
the QME into a set of independent matrix linear equations in a new
matrix unknown $T$ which must be nonsingular. From a practical
point of view the main bottleneck of such a recipe is of course
the determination of  all the distinct roots of $g(\lambda)$, even
if the conditioned resolution of the independent linear system of
$n^2$ equations  to seek the acceptable matrices $T$ might result
quite cumbersome. Details on such a procedure may be found for
example in (Gantmacher 1998), where the method is presented for
right matrix equations of arbitrary degree. Since it is
practically hopeless to give a reply to our questions \textit{Q1)}
and \textit{Q2)} for a generic QME using the above outlined
algorithm, other approaches have been searched  to investigate on
the existence and the number of solutions of a QME.

We quote the following result due to Dennis (1978)
\begin{theorem} If
$g(\lambda)$ has $p$ distinct eigenvalues $\{\lambda_i\}_{i=1}^p$
with $n\leq p\leq 2n$ and the corresponding set of $p$
eigenvectors $\{v_i\}_{i=1}^p$ satisfies the Haar condition, i.e.
every subset of $n$ of them is linearly independent, then the
different solvents of the QME  are exactly $\left(\begin{array}{c}
                                2n\\n
                                \end{array} \right)$, if $p=2n$ and
at least $\left(\begin{array}{c}
                                p\\n
                                \end{array} \right)$, if $n\leq p < 2n$, all by the
                                form
\begin{equation} S=Wdiag(\mu_i)W^{-1}\quad W=[w_1,\dots, w_n]\end{equation}
where the eigenpairs $(\mu_i,w_i)_{i=1}^n$ are chosen from among
the eigenpairs $(\lambda_i,w_i)_{i=1}^p$ of $g(\lambda)$.
\end{theorem}
                                When $p=n$, the distinctness
                                condition of the eigenvalues is
                                not needed, such that we obtain a
                                sufficient condition for the
                                existence of a solvent
                                \begin{corollary}If $g(\lambda)$ has $n$ linearly independent eigenvectors, then QME has a solvent.\end{corollary}
The above method furnishes only  diagonalizable solvents and in
consequence it can fail to identify all solvents and can even
produce no solvents when, in fact, solvents exist.

Some studies on the existence of solvents which  avoid  the
knowledge of the eigensystem $Q(\lambda)$ are known; for example,
based on the contraction mapping principle technique it is
possible to show that if $\mathcal{L}_1$ is a nonsingular matrix
and
\begin{equation}4\|\mathcal{L}_1^{-1}\|\|\mathcal{L}_1^{-1}\mathcal{L}_0\|<1\end{equation}
for any subordinate matrix norm, then at least two solvents exist
( Eisenfeld 1973) (it's worth to note the similarity with the
positivity condition of the discriminant for the QSE case). A
similar but more restrictive condition was derived by McFarland,
Lancaster and Rokne (1977) who, using the Kantorovich theorem on
the convergence of Newton method, have derived several sets of
sufficient conditions for the existence of a solvent. Moreover,
Lancaster 1966, Krein and Lager 1978 have proved that  the so
called "overdamping condition" is sufficient and  ensures the
existence of at least two solvents of a QME, called the dominant
and minimal solvents.

 \vspace{0.2cm}\subsection{\textbf{Solvents
and coefficients }}\vspace{0.2cm}

Putting $\mathcal{L}_1=0$, eq. (\ref{QME})  becomes
 $X^2=\mathcal{L}_0$. Such an equation admits
the resolutive formula $X=\sqrt{\mathcal{L}_0}$ formally similar
to that valid for the QSE, the substantial difference being that
now $\sqrt{\mathcal{L}_0}$  might not exist making accordingly the
equation impossible. \par When $\mathcal{L}_1\neq0$, the question
of the existence of a resolutive formula becomes doubtful. In
fact, it should have the form $X=F(\mathcal{L}_0, \mathcal{L}_1)$,
with $F$ universal and matrix multivalued function of
$\mathcal{L}_0$ and $\mathcal{L}_1$. Since it must reduce to the
well known expression valid in the scalar case, it may be
conjectured that $F$ must contain operations like the square root
of $\Delta\equiv\mathcal{L}_1^2+4\mathcal{L}_0$. Since  this
square root might not exist, one would expect in this case to have
to do with an impossible equation. But, the following example
shows that this argument is generally false. In fact, the equation
$X^2+2\left(\begin{array}{cc}
                                1&0\\0&-1
                                \end{array} \right)X-\left(\begin{array}{cc}
                                -1&1\\0&-1
                                \end{array} \right)=0$
possesses the solvent $\left(\begin{array}{cc}
                                -1&1/2\\0&1
                                \end{array} \right)$
even if the square root of the matrix
$\Delta=\mathcal{L}_1^2+4\mathcal{L}_0=4\left(\begin{array}{cc}
                                0&1\\0&0
                                \end{array} \right)$ does not
exist. \par One might wonder whether for a QME some aspects
characterizing the scalar case such as the commutativity between
$\mathcal{L}_0$ and $\mathcal{L}_1$ and the existence of the
square root of $\Delta=\mathcal{L}^2_1+4\mathcal{L}_0$ are
sufficient
 to represent all the solvents of eq. (\ref{QME}) by
means of a sort of \lq\lq resolutive formula\rq\rq like
$X=(\mathcal{L}_1+\sqrt{\Delta})/2$. The reply is negative due to
the fact that when a specific realization
$\overline{\sqrt{\Delta}}$ of the many-valued symbol
$\sqrt{\Delta}$ is not expressible in the form of a polynomial in
$\Delta$, it might occur that the commutator
$[\mathcal{L}_1,\overline{\sqrt{\Delta}}]\equiv
(\mathcal{L}_1\overline{\sqrt{\Delta}}-\overline{\sqrt{\Delta}}\mathcal{L}_1)$
does not vanish. Thus, the formula
$X=(\mathcal{L}_1+\sqrt{\Delta})/2$ cannot be used as a resolutive
tool even under the two simplifying assumptions, except when, in
addition, $\sqrt{\Delta}$ commutes with $\mathcal{L}_1$, which is
generally not true. Consider indeed the equation
$X^2-\mathcal{L}_1X+\mathcal{L}_1^2/4=0$, where
$\mathcal{L}_0\equiv-\mathcal{L}_1^2/4$ commutes by construction
with $\mathcal{L}_1$ and $\Delta=0$. Of course the square root of
the matrix $\Delta=0$ exists and has infinitely-many realizations
that is the nilpotent matrices of
 index $2$ defining the set $\mathcal{N}_2^{(n)}$. All these matrices, except
the null matrix, cannot of course be expressed in the form of a
polynomial in $\Delta (=0)$  and thus we expect that \lq\lq the
resolutive formula \rq\rq $X=(\mathcal{L}_1+\sqrt{\Delta})/2$ does
not hold even in this simple case. This means that matrix
realizations $\overline{\sqrt{0}}$ of $\sqrt{0}$ not commuting
with $\mathcal{L}_1$ might exist. \\Taking for example
$\mathcal{L}_1=\left(\begin{array}{cc}
                                1&1\\0&0
                                \end{array}
                                \right)=\mathcal{L}_1^2$ and $\overline{\sqrt{0}}=\left(\begin{array}{cc}
                                1&1\\-1&-1
                                \end{array}
                                \right)\in \mathcal{N}_2^{(2)}$
                                where \begin{eqnarray}\mathcal{N}_2^{(2)}\equiv\{N_2^{(2)}= \left(\begin{array}{cc}
                                a&b\\c&-a
                                \end{array}
                                \right)/a,b,c\in\mathbb{C},
                                a^2+bc=0\}\end{eqnarray}
we easily check  that $(\mathcal{L}_1+\overline{\sqrt{0}})/2$ is
not
 a solvent of the equation
                                $X^2-\left(\begin{array}{cc}
                                1&1\\0&0
                                \end{array}
                                \right)X+\frac{1}{4}\left(\begin{array}{cc}
                                1&1\\0&0
                                \end{array}
                                \right)=0$.
                                It is possible to prove that the
                                only solvent of this equation is
                                $\mathcal{L}_1/2$ and that it                                                                                                                                                                                                                                                                                                                                                                                                                                                                                                                                                                                                  is obtainable from
                                $(\mathcal{L}_1+\sqrt{0})/2$
                                solely choosing $\sqrt{0}=0$.
On contrary, the simple QME $(X-E/2)^2=0$ admits infinitely-many
solvents, namely $S=E/2+N_2^{(2)}$, even if once again it has
$\Delta=0$. Thus, for this example, the formula
$X=(\mathcal{L}_1+\sqrt{\Delta})/2$ is valid since
$\mathcal{L}_1=E$ commutes with every realization of
$\sqrt{\Delta}$. Another intriguing example is provided by the
equation $X^2-\mathcal{L}_1X=0$ with
$\mathcal{L}_1=\left(\begin{array}{ccc}
                                0&0&0\\0&0&1\\1/2&0&0
                                \end{array}
                                \right)$ and $\mathcal{L}_0=0$. In
this case $[\mathcal{L}_1,\mathcal{L}_0]=0$,
$\Delta=\mathcal{L}_1^2=\left(\begin{array}{ccc}
                                0&0&0\\1/2&0&0\\0&0&0
                                \end{array}
                                \right)$ and the formula $X=(\mathcal{L}_1+\sqrt{\Delta})/2$
does not include the matrix $\left(\begin{array}{ccc}
                                0&0&0\\0&0&1\\1&0&0
                                \end{array}
                                \right)$ which, in turn, is a
solvent of the equation. This last example is interesting because
it shows that the resolutive formula may fail since it does  not
capture all the solvents of the QME.
\par Summing up, we conjecture heuristically that a general
resolutive formula for QME might not indeed exist. It is however
of relevance to observe that when  $\sqrt{\Delta}$ exists  and
$[\overline{\sqrt{\Delta}},\mathcal{L}_1]=0$, for at least one
realization $\overline{\sqrt{\Delta}}$ of $\sqrt{\Delta}$, then
the QME admits the solvent
$X=(\mathcal{L}_1+\overline{\sqrt{\Delta}})/2$.
\\By the way, it is interesting to note that $\sqrt{\Delta}$ exists when QME (\ref{QME}) admits a
solvent $S$ commuting with $\mathcal{L}_1$. We may indeed write
that
$S^2-\mathcal{L}_1S-\mathcal{L}_0=S^2-\mathcal{L}_1S/2-S\mathcal{L}_1/2+\mathcal{L}_1^2/4-\Delta/4=(S-\mathcal{L}_1/2)^2-\Delta/4=0\Rightarrow$
 $ (S-\mathcal{L}_1/2)^2=\Delta/4$ which, by definition, assures
that $\Delta$ admits square root.

Reversing our point of view, we concentrate now on establishing
whether given two matrices $S_1$ and $S_2\in M_n(\mathbb{C})$,
there always exists at least a QME admitting $S_1$ and $S_2$ as
solvents (eventually, among others solvents). We know that in the
scalar case this important question has a unique positive reply,
that is the knowledge of two roots of a scalar equation determines
the coefficients of the equation (putting legitimately its
director coefficient equal to 1). What can one say about in the
case of QMEs? The reply to this question is more articulated than
in the scalar case in the sense that all the possibilities may now
take place: the given $S_1$ and $S_2$
 may be solvents of only one determinable QME; no QME  could exist
 simultaneously
 admitting the two given  matrices as solvents or there might exist
 infinitely-many QME satisfying such a condition (see Appendix B).
\par In the following a matrix pair $M$, $N\in  M_n(\mathbb{C})$ such
that $\det(M-N)\neq 0$ is called complete, accordingly with the
definition introduced by ref. (Krein and Langer 1978) in the
context of quadratic operator equations.

We now may prove the following
\begin{theorem}Two matrices $S_1$ and $S_2\in M_n(\mathbb{C})$
are simultaneous solvents of only one QME
$X^2-\mathcal{L}_1X-\mathcal{L}_0=0$,
       if and only if the matrix pair $S_1$and $S_2$ is complete  and the
       coefficients $\mathcal{L}_1$ and $\mathcal{L}_0$ may be
       expressed as
\begin{equation}\label{EXL0}\mathcal{L}_0=S_2^2(S_1-S_2)^{-1}S_1-S_1^2(S_1-S_2)^{-1}S_2\end{equation}

\begin{equation}\label{EXL1}\mathcal{L}_1=(S_1^2-S_2^2)(S_1-S_2)^{-1}\end{equation}

\end{theorem}
\emph{Proof}:\par The existence of a unique QME admitting $S_1$
and $S_2$ as simultaneous solvents of eq. (\ref{QME}) requires
that

\begin{equation}\left\{\begin{array}{rl}
  S_1^2-\mathcal{L}_1S_1-\mathcal{L}_0=0\\
     S_2^2-\mathcal{L}_1S_2-\mathcal{L}_0=0  \end{array}\right.\end{equation}
or equivalently that
\begin{equation}\left\{\begin{array}{rl}
 \mathcal{L}_0=S_1^2-\mathcal{L}_1S_1=S_2^2-\mathcal{L}_2S_2\\
     S_1^2-S_2^2= \mathcal{L}_1(S_1-S_2)\end{array}\right.\end{equation}
Regarding the second equation as a matrix linear equation in
$\mathcal{L}_1$ we know by hypothesis that it has only one
solution. This fact necessarily requires that $rank(S_1-S_2)=n$,
that is $(S_1-S_2)$ is not singular and thus necessarily that
\begin{equation}\label{SOSS}\mathcal{L}_1=(S_1^2-S_2^2)(S_1-S_2)^{-1}\end{equation}
To determine $\mathcal{L}_0$ in the desired form we observe that
\begin{center}$
\mathcal{L}_0=S_1^2-(S_1^2-S_2^2)(S_1-S_2)^{-1}S_1=S_1^2(S_1-S_2)^{-1}(S_1-S_2)-(S_1^2-S_2^2)(S_1-S_2)^{-1}S_1$\end{center}
\begin{equation}
\label{SOS}
=S_2^2(S_1-S_2)^{-1}S_1-S_1^2(S_1-S_2)^{-1}S_2\end{equation}

\par Reciprocally, if the pair $S_1$ and $S_2$ is complete and $\mathcal{L}_0$
and $\mathcal{L}_1$ are given by eqs. (\ref{EXL0}) and
(\ref{EXL1}) respectively, then it is immediate to see that $S_1$
and $S_2 $ are solvents of the corresponding QME. \par If
$(S_1-S_2)$ is not invertible, then  the necessary condition
$S_1^2-S_2^2= \mathcal{L}_1(S_1-S_2)$ may be incompatible with the
existence of a QME having $S_1$ and $S_2$ as solvents or may give
rise to infinitely-many possibilities for $\mathcal{L}_1$ and
$\mathcal{L}_0$. This means that there might even exist infinitely
many QME possessing $S_1$ and $S_2$ as simultaneous solvents. In
the appendix B we give an example of  two matrices $S_1$ and $S_2$
which cannot be simultaneous solvents of any QME and two matrices
$S'_1$ and $S'_2$ which are simultaneous solvents of infinitely
many QME. \\ We underline that eqs. \eqref{EXL0} and \eqref{EXL1}
may be indirectly constructed exploiting  lemma 2.14 of
ref.Gohberg 1982.

 \vspace{0.2cm} \section{Symmetric
functions associated to a QME}\vspace{0.2cm}
\subsection{Elementary symmetric functions associated to a QME}
\par It is interesting to point out that under the condition of
validity of eqs. (\ref{EXL0}) and
 (\ref{EXL1}), the further assumption that $[S_1, S_2]=0$ leads us,
 as
 expected, to the well known elementary relations
 $\mathcal{L}_0=-S_1S_2$ and $\mathcal{L}_1=S_1+S_2$ which in turn implies that $[\mathcal{L}_0, \mathcal{L}_1]=0$.
 Even if we are sure that, being in addition $\Delta=(S_1-S_2)^2$,
 $\sqrt{\Delta}$ exists, we do not fully recover all the properties
 holding in the scalar case since, as previously noted, there might exist matrix values of $\sqrt{\Delta}$
not commuting with $\mathcal{L}_1$. The two expressions $S_1+S_2$
and $S_1S_2$ play an important role in the context of the theory
of quadratic scalar equations. They indeed are the two associated
elementary symmetric functions of the two (only and always
existing) roots of the QSE, in terms of which every rational
symmetric function of the same roots $S_1$ and $S_2$ is rationally
expressible. It is immediate to persuade ourselves that for a QME
admitting at least the two solvents $S_1$ and $S_2$, matrix
functions like $S_1+S_2$ and $S_1S_2$, although the simplest ones
possessing permutational invariance under the exchange of $S_1$
with $S_2$, cannot occupy a special place for the QME simply
because they are not invariant in the set $\mathcal{S}$ of all the
possible solvents of the QME. This means that in general
$S_1+S_2\neq S_1'+S_2'$ and/or $S_1S_2\neq S_1'S_2'$ with
$S_1,S_2,S_1',S_2'\in \mathcal{S}$ so that  these functions are
not biunivocally determined by the coefficients of the QME under
scrutiny. Theorem 1 suggests to identify the right hand sides of
eq. (\ref{EXL0}) and (\ref{EXL1}) as the elementary symmetric
functions of two solvents to be used when dealing with a QME.
These expressions indeed are
  not only permutationally invariant but their matrix values
do not change when we substitute a complete pair of solvents
$S_1$, $S_2$ with another one complete too. For this reason we are
stimulated to introduce the following
\begin{definition}
Let  $S_1$  and $S_2$ be a complete pair of solvents of a QME
(\ref{QME}).\\ The two expressions
\begin{equation}\label{sum}\Sigma_2(S_1,S_2)=(S_1^2-S_2^2)(S_1-S_2)^{-1}\end{equation}
and
\begin{equation}\label{prod}\Pi_2(S_1,S_2)=S_1^2(S_1-S_2)^{-1}S_2-S_2^2(S_1-S_2)^{-1}S_1\end{equation}
are called elementary symmetric functions associated to a QME.
\end{definition}
\subsection{Symmetric functions associated to a QME}
It is noteworthy to observe that for a given QME, simple
permutationally invariant expressions like
$\mathbf{S}_{p,2}=S_1^p+S_2^p$ or
$\mathbf{\Pi}_{p,2}=S_1^pS_2+S_2^pS_1$ with $p=1,2,\dots,$
generally lacking of invariance in $\mathcal{S}$ whatever $p$ is,
 cannot be in general expressed in terms of
$\mathcal{L}_0$ and $\mathcal{L}_1$.  We, thus, claim that it is
not true that every symmetric function of two solvents of a QME
not possessing permutational invariance in $\mathcal{S}$, is
rationally expressible in terms of the elementary symmetric
functions associated to a QME (\ref{sum}) and (\ref{prod}) only.
In particular, this means that even $\mathbf{S}_{1,2}=S_1+S_2$ and
$\mathbf{\Pi}_{1,2}=S_1S_2+S_2S_1$ suffer such a difficulty and
for this reason  we call them irreducible symmetric functions. We
are however going to show that when $S_1$ and $S_2$ constitute a
complete pair of solvents of a QME, then $\mathbf{S}_{p,2}$ and
$\mathbf{\Pi}_{p,2}$ may be represented in terms of the elementary
symmetric functions $\Sigma_2(S_1,S_2)$ and $\Pi_2(S_1,S_2)$ given
by eqs. (\ref{sum}) and (\ref{prod}) respectively as well as in
terms of the irreducible expressions $\mathbf{S}_{1,2}$ and
$\mathbf{\Pi}_{1,2}$ respectively. We anticipate that such a
representation of $S_{p,2}$ may be interpreted as a generalization
of the  Waring formula, well-known in the context of QSE.
 \par  To this end
we exploit a recently published result concerning the exact
resolution of the matrix Cauchy problem
\begin{equation}\label{CP}\left\{\begin{array}{rl}
      Y_{p+2}=\mathcal{L}_0Y_p+\mathcal{L}_1Y_{p+1}\\
       Y_0=0,\quad Y_1=B\end{array}\right.,\end{equation}
where the variable $p$ runs in $ \mathbb{N}$ and $Y_p$,
$\mathcal{L}_1$ and $\mathcal{L}_0$ belong to $M_n(\mathbb{C})$
(Jivulescu et al. 2007). We have been able to extend the
successful technique envisaged in this reference to solve the more
general Cauchy problem
\begin{equation}\label{CPP}\left\{\begin{array}{rl}
      Y_{p+2}=\mathcal{L}_0Y_p+\mathcal{L}_1Y_{p+1}\\
       Y_0=A,\quad Y_1=B\end{array}\right., \end{equation}
getting the following always existing unique matrix solution
\begin{equation} \label{S}Y_p=\alpha_p A+ \beta_p B, \end{equation}
where
\begin{equation}\label{alfa} \alpha_p=\left\{\begin{array}{ll}
\sum\limits_{t=0}^{[\frac{p-2}{2}]}\{\mathcal{L}_0^{(t)}\mathcal{L}_1^{(p-2-2t)}\}\mathcal{L}_0&\quad if \quad p\geq2 \\
 0&\quad if \quad  p=1\\
E&\quad if \quad p=0
\end{array}\right.\end{equation}and
\begin{equation}\label{beta}
\beta_p=\left\{\begin{array}{ll}\sum\limits_{t=0}^{[\frac{p-1}{2}]}\{\mathcal{L}_0^{(t)}\mathcal{L}_1^{(p-1-2t)}\}
& \quad if  \quad p\geq1\\0& \quad if \quad p=0
\end{array}\right.,
\end{equation}
where $E$ is the unitary matrix. The  symbol $\{\mathcal{L}_0
^{(u)}\mathcal{L}_1^{(v)}\}$, introduced in (Jivulescu et al.
2007), denotes the sum of all possible distinct permutations of
$u$ factors $\mathcal{L}_0$ and $v$ factors
 $\mathcal{L}_1$.
 \par As an immediate application of eqs. (\ref{CPP}) and (\ref{S}), let
firstly notice that if $S$ is a solvent of eq. (\ref{QME}), then
$S^2-\mathcal{L}_1S-\mathcal{L}_0=0$ and consequently
$S^{p+2}-\mathcal{L}_1S^{p+1}-\mathcal{L}_0S^p=0$, where
$p\in\mathbb{N}$. This means that the $p-$th power of any
arbitrary solvent $S$ of eq. (\ref{QME}) satisfies the following
Cauchy problem
\begin{equation}\label{CP}\left\{\begin{array}{ll}
      S^{p+2}=\mathcal{L}_0S^p+\mathcal{L}_1S^{p+1}\\
       S^0=E,\quad S^1=S\end{array}\right. \end{equation}
which, coinciding with that expressed by eq. (\ref{CPP}), leads to
the following formula
\begin{equation}\label{POW}S^p=\beta_pS+\alpha_pE,\end{equation}
 \par It is worth noticing that the two
operator coefficients $\alpha_p$ and $\beta_p$ appearing in the
right hand side of eq. (\ref{POW}) are completely determined by
the knowledge of the coefficients $\mathcal{L}_0$ and
$\mathcal{L}_1$ only. Thus, eq. (\ref{POW}) can be regarded as an
effective linearization of $S^p$, for any solvent $S$, becoming
indeed a computational tool in the sense that the expressions
$\alpha_p$ and $\beta_p$, for a given $p$, may be calculated only
once, whatever the solvent $S$ is.

In passing we observe that if $S$ represents the transfer matrix
of a \lq\lq box"  then, whatever its nature is, $S^p$ may be
interpreted as the output after $p$ identical iterations, the QME
admitting $S$ as solvent implicitly defining the transmission
features of the box. In the last section we will exploit this
point of view to deduce the transfer matrix of an $N$-period
potential.
\par Eq. (\ref{POW}) can be generalized. Given indeed $r$-simultaneous solvents
$S_1,S_2,\dots,S_r$ of eq. (\ref{QME}), let's introduce
\begin{equation}\label{Sum}\mathbf{S}_{p,r}=\sum\limits_{i=1}^rS_i^p\end{equation}
a permutationally invariant quantity  under the exchange of any
$i$ with any $j$, $(i,j=1,\dots, r)$. It is immediate to persuade
oneself that such a quantity satisfies the Cauchy problem
\begin{equation}\label{CPPP}\left\{\begin{array}{rl}
      \mathbf{S}_{p+2,r}=\mathcal{L}_0\mathbf{S}_{p,r}+\mathcal{L}_1\mathbf{S}_{p+1,r}\\
       \mathbf{S}_{0,r}=rE,\quad \mathbf{S}_{1,r}=\sum\limits_{i=1}^rS_i\end{array}\right.\end{equation}
which, in view of  eq. (\ref{S}), admits the solution
\begin{equation}\label{Y}\mathbf{S}_{p,r}=\beta_p\left(\sum\limits_{i=1}^rS_i\right)+r\alpha_pE\end{equation}
This formula is at glance deducible from eq. (\ref{POW}) which in
turn is recoverable from eq. (\ref{Y}) simply putting $r=1$. We
emphasize that arbitrarily fixing $p$, the expression of
$\mathbf{S}_{p,r}$ given by eq. (\ref{Y}) is linearly related to
$\mathbf{S}_{1,r}$ in the sense that the operator coefficients
$\alpha_p$ and $\beta_p$ are independent both on the number $r$ of
solvents involved and on the group of $r$- solvents chosen in the
set $\mathcal{S}$ of all solvents of eq. (\ref{QME}).

We recall that the set $\mathcal{S}$ may be also infinite thus
permitting us to chose in infinitely many ways the family of $r$-
solvents satisfying eq. (\ref{Y}). In addition, like eq.
(\ref{POW}), this equation may be regarded as a computational tool
for the $p$-power sum of $r$ solvents of eq. (\ref{QME}).

 \par We wish to notice that putting $p=r=2$ into eq. (\ref{Y}), under the condition that $S_1$ and $S_2$ are a complete pair of simultaneous solvents of eq. (\ref{QME}), yields
\begin{equation}\label{GN}\mathbf{S}_{2,2}=\Sigma_2(S_1,S_2)\mathbf{S}_{1,2}-2\Pi_2(S_1,S_2)E\end{equation}
 which may be interpreted as generalization to the matrix case
 of the
classical Girard- Newton formula for symmetric polynomials in two
scalar variables (Sansone 1952).
 \par In the same spirit and hypothesis we are also able to get an  extension to the matrix case of the well-known Waring formula
for two variables. Putting indeed $r=2$ into the identity
(\ref{Y}), leaving instead the integer $p$ free to run in
$\mathbb{N}$, we immediately get
\begin{equation}\label{SP}\mathbf{S}_{p,2}=\beta_p(S_1+S_2)+2\alpha_pE\end{equation}
\par The peculiar feature of eq. (\ref{SP}) is that the
presence of the operators $\alpha_p$ and $\beta_p$ amounts at
rewriting the $p$-th power sum $\mathbf{S}_{p,2}$ in terms of
$\Sigma_2(S_1,S_2),\Pi_2(S_1,S_2)$, the above defined elementary
symmetric functions associated to the QME, for this reason we
claim that eq. (\ref{SP}) may be interpreted as generalization to
the matrix case of the Waring formula for two scalar variables
(Sansone 1952) .
\par Another direct consequence of eq. (\ref{CPP}) concerns the
permutationally invariant quantity
\begin{equation}\label{Prod}\mathbf{\Pi}_{p,r}=\sum\limits_{1\leq
i<j\leq r}(S_i^pS_j+S_iS_j^p)\end{equation}

 It is not difficult to see that  the quantity
$(S_i^pS_j+S_j^pS_i)$, for any given $i$ and $j$, satisfies eq.
(\ref{POW}),  that is
\begin{equation}S_i^pS_j+S_j^pS_i=\beta_p(S_iS_j+S_jS_i)+\alpha_p(S_i+S_j)\end{equation}
 from
 which we immediately deduce that
\begin{equation}\label{R}\mathbf{\Pi}_{p,r}=\beta_p\mathbf{\Pi}_{1,r}+(r-1)\alpha_p\mathbf{S}_{1,r}\end{equation}

\par
Applying to eq. (\ref{R}) the same arguments used to deduce eq.
(\ref{SP}) from eq. (\ref{Y}) immediately yields
\begin{equation}\label{PI}\mathbf{\Pi}_{p,2}=\alpha_p\mathbf{S}_{1,2}+\beta_p\mathbf{\Pi}_{1,2}\end{equation}
which is of relevance since it expresses $\mathbf{\Pi}_{p,2}$
 in terms of the irreducible expressions $\mathbf{S}_{1,2}=(S_1+S_2)$ and $\mathbf{\Pi}_{1,2}=S_1S_2+S_2S_1$ and
the elementary symmetric functions associated to the QME
$\Sigma_2(S_1,S_2),\Pi_2(S_1,S_2)$, considering that $\alpha_p$
and $\beta_p$ are functions of $\mathcal{L}_0=\Sigma_2(S_1,S_2)$
and $\mathcal{L}_1=-\Pi_2(S_1,S_2)$. \par It is interesting to
stress that the unavoidable presence of $\mathbf{S}_{1,2}$ in eq.
(\ref{SP}) and of $\mathbf{\Pi}_{1,2}$ too into eq. (\ref{PI})
does not allow to write down the permutationally invariant
expressions $\mathbf{S}_{p,2}$ and $\mathbf{\Pi}_{p,2}$ in terms
of the noncommutative coefficients $\mathcal{L}_0$ and
$\mathcal{L}_1$ only. This observation confirms our claim that the
fundamental theorem regarding the symmetric functions of two
scalar variables
 does not hold in the matrix case. \par The elementary symmetric
functions
 associated to the QME (\ref{QME}) may be easily generalized.
To this end let's write  eq. (\ref{POW}) for both elements $S_1$
and $S_2$ of a complete pair of simultaneous solvents
\begin{equation}\left\{\begin{array}{rl}
  S_1^p=\beta_pS_1+\alpha_p\\
     S_2^p=\beta_pS_2+\alpha_p  \end{array}\right.\end{equation}
It is simple to solve this system of equations in the unknowns
$\alpha_p$ and $\beta_p$ obtaining
\begin{equation}\label{III}\alpha_p=S_2^p(S_1-S_2)^{-1}S_1-S_1^p(S_1-S_2)^{-1}S_2\end{equation}
and
\begin{equation}\label{IIII}\beta_p=(S_1^p-S_2^p)(S_1-S_2)^{-1}\end{equation}
Eqs. (\ref{III}) and (\ref{IIII}) reduce to $-\Pi_2(S_1, S_2)$ and
$\Sigma_2(S_1,S_2)$ when $p=2$. In addition, the right hand side
of both equations are certainly invariant in the set
$\mathcal{S}$. For this reason, we are induced to introduce the
following

\begin{definition}
Let  $S_1$  and $S_2$ be a complete pair of simultaneous solvents
of a QME (\ref{QME}). The two expressions
\begin{equation}\label{sig}\Sigma_p(S_1,S_2)=(S_1^p-S_2^p)(S_1-S_2)^{-1}\end{equation}
and \begin{equation}\label{prod}
\Pi_p(S_1,S_2)=S_1^p(S_1-S_2)^{-1}S_2-S_2^p(S_1-S_2)^{-1}S_1\end{equation}
are called symmetric functions associated to a QME.
\end{definition}
It is of relevance that $\Sigma_p(S_1,S_2)$ and $\Pi_p(S_1,S_2)$
may be expressed in terms of elementary symmetric functions
associated to a QME $\Sigma_2(S_1,S_2)$ and $\Pi_2(S_1,S_2)$  as
easily seen considering the definition of $\alpha_p$ and $\beta_p$
and in the view of the eqs. (\ref{EXL0}) and (\ref{EXL1}). If we
restrict our definition of symmetric functions associated to a QME
to all those expressions of two complete solvents satisfying the
condition to be permutationally invariant into $\mathcal{S}$ then
such expressions like \eqref{sig} and \eqref{prod} must be
expressible in terms of $\mathcal{L}_0$ and $\mathcal{L}_1$  and
then in terms of $\Sigma_2(S_1,S_2)$ and $\Pi_2(S_1,S_2)$.
 \par Thus, we claim that any symmetric function associated to a
QME may be written down in terms of elementary symmetric
functions.
\section{A simple application}
\par If the order of the unknown matrix $X$ of eq. (\ref{QME})
is fixed at its minimum value $1$ we recover the scalar case
getting simultaneously rid of both the matrix nature of the
problem and of any complication stemming from the noncommutativity
between $\mathcal{L}_0$ and $\mathcal{L}_1$ as well as between $X$
and the two coefficients. A simpler but not trivial problem,
someway intermediate between the full matrix noncommutative eq.
(\ref{QME}) and its scalar version, may be singled out requiring
solely the commutativity between the two coefficients
$\mathcal{L}_0$ and $\mathcal{L}_1$. We emphasize that such a
restriction does not spoil eq. (\ref{QME}) of interest enabling
us, in turn, to bring to light some peculiar properties consequent
to its matrix nature only. We stress indeed that on the basis of
the results reported in the previous sections of this paper, the
hypothesis $[\mathcal{L}_0,\mathcal{L}_1]=0$ does not lead to any
effective recipe to build up the set of solvents of eq.
(\ref{QME}) which may be still empty or contain finitely-many or
infinitely many elements. The concrete advantage provided by the
condition $[\mathcal{L}_0, \mathcal{L}_1]=0$ is that it makes the
matrix expressions of $\alpha_p$ and $\beta_p$ algebraically
manageable disclosing their connection with the
 Chebyshev polynomials of the second kind
$\mathcal{U}_p(x)$ (Murray-Spiegel 1998) defined as follows
\begin{equation}\mathcal{U}_p[x]=\sum\limits_{m=0}^{[p/2]}(-1)^m\frac{(p-m)!}{m!(p-2m)!}(2x)^{p-2m}\end{equation}
Taking indeed into consideration the commutativity between
$\mathcal{L}_0$ and $\mathcal{L}_1$ as well as the fact that the
number of all the different terms appearing in the operator symbol
$\{\mathcal{L}_0 ^{(u)}\mathcal{L}_1^{(v)}\}$ coincides with the
binomial coefficient $ \left(\begin{array}{c}
                                u+v\\ m
                                \end{array} \right)$, with $m=\min(u,v)$, we may
rewrite for $p\geq 2$ that
\begin{equation}\alpha_p=\sum\limits_{t=0}^{[(p-2)/2]} \left(\begin{array}{c}
                                p-2-t\\ t
                                \end{array} \right)\mathcal{L}_0^{t+1}\mathcal{L}_1^{p-2-2t}\end{equation}
while $\alpha_0=E$ and $\alpha_1=0$. Assuming the existence of the
inverse of the square root of $\mathcal{L}_0$ opens the way to
express $\alpha_p$ as\begin{eqnarray}
\alpha_p=-(-\mathcal{L}_0)^{p/2}
\sum\limits_{t=0}^{[(p-2)/2]}(-1)^t\left(\begin{array}{c}
                                p-2-t\\ t
                                \end{array} \right)(\mathcal{L}_1(-\mathcal{L}_0)^{-\frac{1}{2}})^{(p-2-2t)}\end{eqnarray} that is
\begin{equation}\label{C1}\alpha_p=-(-\mathcal{L}_0)^{\frac{p}{2}}\mathcal{U}_{p-2}\left[\frac{1}{2}\mathcal{L}_1(-\mathcal{L}_0)^{-\frac{1}{2}}\right],\quad p\geq 2\end{equation}
Following a similar procedure and under the same hypothesis we
have
\begin{center}$\sum\limits_{t=0}^{[(p-1)/2]}\left(\begin{array}{c}
                                p-1-t\\ t
                                \end{array} \right)\mathcal{L}_0^t\mathcal{L}_1^{p-1-2t}
=(-\mathcal{L}_0)^{(p-1)/2}\sum\limits_{t=0}^{[(p-1)/2]}(-1)^t\left(\begin{array}{c}
                                p-1-t\\ t
                               \end{array} \right)(\mathcal{L}_1(-\mathcal{L}_0)^{-\frac{1}{2}})^{(p-1-2t)}$\end{center}
from which
\begin{equation}\label{C2}\beta_p=(-\mathcal{L}_0)^{\frac{p-1}{2}}\mathcal{U}_{p-1}\left[\frac{1}{2}\mathcal{L}_1(-\mathcal{L}_0)^{-\frac{1}{2}}\right], \quad p\geq 1\end{equation}
while $\beta_0=0$. Inserting eqs. (\ref{C1}) and (\ref{C2}) into
eq. (\ref{POW}) yields
\begin{equation}
\label{POWC}S^p=(-\mathcal{L}_0)^{\frac{p-1}{2}}\mathcal{U}_{p-1}\left[\frac{1}{2}\mathcal{L}_1(-\mathcal{L}_0)^{-\frac{1}{2}}\right]S-
(-\mathcal{L}_0)^{\frac{p}{2}}\mathcal{U}_{p-2}\left[\frac{1}{2}\mathcal{L}_1(-\mathcal{L}_0)^{-\frac{1}{2}}\right]E,\quad
p \geq 2\end{equation} which reduces to the trivial identities
$S=S$ and $E=E$, when $p=1$ and $p=0$ respectively. \\If the two
coefficients $\mathcal{L}_0$ and $\mathcal{L}_1$ appearing in eq.
(\ref{POW}) commute, then the right hand side of formula
(\ref{POWC}) provides the definition of the transfer matrix of a
single \lq\lq equivalent box " substituting all the original $p$
ones.\par It is worth noting that the practical use of eq.
(\ref{POWC}) is greatly facilitated when the minimal polynomial of
$S$ has a degree $2$. We remind that the monic polynomial of
minimum degree $p(x)$ such that $p(S)$ is the zero matrix is
called the minimal polynomial of $S$ and that it always exists and
is unique (Gantmacher 1998). Let us suppose that the distinct
roots of the characteristic polynomial of $S$, a given solvent of
eq. (\ref{QME}), are only two and denote them by $\lambda_1$ and
$\lambda_2$. If in addition we stipulate that $S$ is
diagonalizable, it is well-known that, then, its minimal
polynomial coincides with the characteristics polynomial of the
auxiliary two by two diagonal matrix
$\widetilde{S}=\left(\begin{array}{cc}
                                \lambda_1&0\\0&\lambda_2
                                \end{array}
                                \right)$.
This amounts at claiming that $S$ is also solvent of the QME
$S^2-(tr \widetilde{S}) S+(det\widetilde{S})E=0$, relatively
simpler than eq. (\ref{QME}) since $\mathcal{L}_0$ and
$\mathcal{L}_1$ are now proportional to $E$. As a direct
consequence formula (\ref{POWC}) may be easily exploited to find
$S^p$, for any $p$. We underline that, when $S$ is of order two,
the applications of this formula may be made systematically
easier, simply having recourse to the Cayley-Hamilton identity
$S^2-(tr S) S+(detS)E=0$. Quite recently, for example, this
approach has been used to express the $N$- period transfer matrix
$M^N$ of a photonic crystal as a linear function of the unit
cell-matrix $M=\left(\begin{array}{cc}
                                1/t&r^*/t^*\\r/t&1/t^*
                                \end{array}
                                \right)$, where $t$ and $r$ are the complex transmission and reflection  coefficients of the unit cell
respectively, satisfying the condition $|r|^2+|t|^2=1$ (Benickson
et al. 1996). It is easy to persuade ourselves that
 the Cayley-Hamilton theorem leads in such a case to the following simple QME
\begin{center}$M^2-2\cos\beta M+E=0$\end{center}
where $\cos \beta=Re(1/t)$.\\ Thus, resorting to eq. (\ref{POWC}),
we may soon establish the transfer-matrix reduction identity
\begin{center}$M^N=\mathcal{U}_{N-1}[\cos\beta]M-\mathcal{U}_{N-2}[\cos\beta]E$\end{center}
which may be put into the following closed form (Murray-Spiegel
1998)
\begin{center}$M^N=\frac{\sin N\beta}{\sin \beta}M-\frac{\sin
(N-1)\beta}{\sin \beta}E$\end{center} This expression coincides
with that differently derived in ref. (Bendickson et al. 1996).
\section{Conclusive remarks}
\par In this paper we have studied the quadratic matrix equation
$X^2-\mathcal{L}_1X-\mathcal{L}_0=0$ bringing to light new
peculiar properties of its, clearly traceable back to the inherent
noncommutative character of the problem. Driven by the knowledge
of any aspect concerning the simple classical theory of the scalar
quadratic equations, we have formulated and coped with some basic
questions putting into evidence remarkable differences between the
matrix and the scalar case. Exploring the meaning and the
consequences of such differences we have succeeded in introducing
the definition and the explicit expressions of the matrix
elementary symmetric functions associated to a QME by which some
notable identities involving solvents of the QME have been
derived. \par Other questions greatly stimulate our attention like
for example the notion of multiple solvents or how to define a
factorization procedure, if any, of a quadratic matrix polynomial
from the knowledge of its finitely-many or infinitely-many
solvents. We believe that to develop the theory of the matrix
algebraic equations, quadratic or of higher order, might provide
useful tools to overcome technical mathematical obstacles
encountered in the resolution of physical problems both in a
classical and in a quantum context.
\subsection{Acknowledgments} The
authors would like to thank  professor A. Giambruno, dr. A.
Settimi, dr. B. Militello and dr. R. Messina for fruitful
discussions.
\appendix{}
\section{APPENDIX}
  Let's give few
examples  clearly illustrating our affirmation that the
fundamental theorem of algebra does not hold for matrix equation;
thus, our examples point out QMEs having no solvent or infinitely
many solvents as well as equations with one or finitely-many
solvents. In fact, these different scenarios are mainly traceable
back to the existence of the square root of a $n\times n$ matrix,
say $M$, which is a multivalued operation generally not
representable in the form of a polynomial in $M$ (Horn 1999).
\par Firstly, let's
consider the quadratic matrix equation $X^2-2EX-\mathcal{L}_0=0$,
where $\mathcal{L}_0$ is the square matrix of order $n$ with its
diagonal entries equal to $-1$, the elements of the first
superdiagonal equal to unit and all the other vanishing. This
equation may be immediately written as $(X-E)^2=\mathcal{L}_0+E$
and it does not admit solvents  since the square roots of
$\mathcal{L}_0+E\equiv H^{(n)}$ does not exist (Gantmacher 1998).
\par The matrix equation $X^2+EX=\mathcal{L}_0$ with
$\mathcal{L}_0$ $n\times n$ diagonal matrix having
$(\mathcal{L}_0)_{ii}=2\delta_{i,1}$ admits finitely-many diagonal
solvents in the form
\begin{equation}\label{1}S_1=\{1,\epsilon_2,\epsilon_3,\dots,\epsilon_n\}, \quad S_{-2}=\{-2,\nu_2,\nu_3,\dots,\nu_n\}\end{equation}
where each of the diagonal entries
$\epsilon_2,\epsilon_3,\dots,\epsilon_n$ in $S_1$ and
$\nu_2,\nu_3,\dots,\nu_n$ in $S_{-2}$ may arbitrarily assume the
values $0$ or $-1$. \par Finally, the matrix equation $X^2-EX=2E$
provides an example wherein infinitely-many solvents exist having
the form of lower triangular matrices
\begin{eqnarray}\label{2}X=\left(\begin{array}{ccccccc}
                                -1&\\c_1&2
                                \\ 0&0&-1
                                \\c_2&&&2&
                                \\\dots
                                \\c_{n/2}&&&&&-1
                                \\0&&&&&&2
                                \end{array}
                                \right)_{n-even},\nonumber \\ X=\left(\begin{array}{ccccccc}
                                -1\\c_1&2
                                \\ 0&0&-1
                                \\c_2&&&2
                                \\\dots
                                \\c_{n-1/2}&&&&&2
                                \\0&&&&&&-1
                                \end{array}
                                \right)_{n-odd}\end{eqnarray} where all the non-indicated entries are equal to $0$
and
$c_1,c_2,\dots,c_{\frac{n}{2}}(c_{\frac{n-1}{2}})\in\mathbb{C}$
  \vspace{0.5cm}

\section{Appendix}The $n$-quadratic matrices
\begin{eqnarray}S_1=\left(\begin{array}{ccccc}
                                1&2&2&\dots&2\\0&1&2&\dots &2
                                \\ \dots&\dots&\dots&\dots &\dots
                                \\0&0&0&\dots &1
                                \end{array}
                                \right),
                                S_2=\left(\begin{array}{ccccc}
                                -1&0&0&\dots&0\\-2&-1&0&\dots &0
                                \\ \dots&\dots&\dots&\dots &\dots
                                \\-2&-2&-2&\dots &-1
                                \end{array}
                                \right)\end{eqnarray}cannot be both solvents
                                of the same QME due to fact that the matrix
                                equation which furnishes the
                                coefficient $\mathcal{L}_1$
\begin{equation}\mathcal{L}_1(S_1-S_2)=(S_1^2-S_2^2)\end{equation}
admits no solution. The incompatibility of equation is based on
the Rouche-Capelli's theorem because $rank(S_1-S_2)\neq
rank(S_1^2-S_2^2)$, where
\begin{eqnarray} S_1-S_2=\left(\begin{array}{ccccc}
                                2&2&2&\dots&2\\2&2&2&\dots &2
                                \\ \dots&\dots&\dots&\dots &\dots
                                \\2&2&2&\dots &2
                                \end{array}
                                \right),\\\nonumber
 S_1^2-S_2^2=\left(\begin{array}{ccccc}
                                0&1\cdot 2^2&2\cdot2^2&\dots&(n-1)\cdot2^2\\-1\cdot 2^2&0&1\cdot 2^2&\dots &(n-2)\cdot
                                2^2
                                \\ \dots&\dots&\dots&\dots &\dots
                                \\-(n-1)\cdot 2^2&-(n-2)\cdot2^2&-(n-3)\cdot 2^2&\dots
                                &0
                                \end{array}
                                \right)\end{eqnarray}
\par
Further, the $n$- quadratic matrices
\begin{equation}S_1'=\left(\begin{array}{ccccc}
                                1&0&...&0&0\\0&1&...&0&0\\...&...&...&...&...\\0&0&...&1&0\\0&0&...&0&3
                                \end{array}
                                \right), S_2'=\left(\begin{array}{ccccc}
                                1&0&...&0&0\\0&1&...&0&0\\...&...&...&...&...\\0&0&...&1&0\\0&0&...&0&2
                                \end{array}
                                \right)\end{equation}
 are two solvents of the class of the QME for which
\begin{eqnarray}\mathcal{L}_1=\left(\begin{array}{ccccc}
                                a_{11}&a_{12}&...&a_{1n-1}&0\\a_{21}&a_{22}&...&a_{2n-1}&0\\...&...&...&...&...\\a_{n-11}&a_{n-12}&...&a_{n-1n-1}&0\\a_{n1}&a_{n2}&...&a_{nn-1}&5
                                \end{array}
                                \right),\nonumber\\ \mathcal{L}_0=\left(\begin{array}{ccccc}
                                1-a_{11}&-a_{12}&...&-a_{1n-1}&0\\-a_{21}&1-a_{22}&...&-a_{2n-1}&0\\...&...&...&...&...\\-a_{n-11}&-a_{n-12}&...&1-a_{n-1n-1}&0\\-a_{n1}&-a_{n2}&...&-a_{nn-1}&-6
                                \end{array}
                                \right),a_{ij}\in\mathbb{C}\end{eqnarray}

\end{document}